\documentstyle[prd,aps,floats,epsfig,graphicx]{revtex}
\input epsf
\def\pht{\tilde{\phi}}
\def\xt{\tilde{x}}
\def\pt{\partial}
\def\s{\sigma}
\newcommand{\be}{\begin{equation}}
\newcommand{\ee}{\end{equation}}
\newcommand{\bea}{\begin{eqnarray}}
\newcommand{\eea}{\end{eqnarray}}
\begin{document}

\title{\bf{Near-Extremal Black Hole Evaporation in Asymptotically Flat Spacetime}}
\author{Kamran Diba\thanks{
e-mail: diba@het.brown.edu} and David A. Lowe}
\address{Department of Physics, \\Brown University,
\\Providence,
RI 02912, \\U.S.A.}

\maketitle

\begin{abstract}
We study black hole evaporation of near-extremal black holes in
spherically reduced models with asymptotically Minkowskian
spacetime, with the effects of the back-reaction on the geometry
included semi-classically. The stress-energy tensor is calculated
for null in-falling observers. It is shown that the evaporation
proceeds smoothly and there are no instabilities of the outer or
inner apparent horizon before the endpoint of evaporation.
\end{abstract}

\section{Introduction}
The evolution of near-extremal black holes has important
consequences for the resolution of the information paradox.
Extremal black holes are potential candidates for endpoint
remnants \cite{gidinfo1,gidinfo2}, and can potentially store
information behind the double horizon \cite{bols}.  In addition,
study of their evaporation avoids the naked singularities observed
in models of linear dilaton black hole evaporation
\cite{cghs,lowefirst}, which are likely typical of the evolution of
uncharged black holes..

However, it has been suggested by Jacobson, using adiabatic
arguments, that the semi-classical evolution of near-extremal
black holes may break down while still far from extremality
\cite{jacobson}.  The claim is that in-falling photons created at
the outer apparent horizon during Hawking evaporation will
eventually fall through the inner horizon and either encounter a
large buildup of energy behind the inner horizon, or
alternatively, pile up behind the outer horizon, causing it to
become unstable.  Thus, in either scenario, the semiclassical
approximation may break down long before one would expect based on
thermodynamic/statistical mechanics arguments \cite{wilczetc}.

This claim has been explored in an analytical model
\cite{dibalowe} using a near-horizon Reissner-Norsdr\"om model
originally proposed in \cite{nav1,nav2}, and further elaborated in 
\cite{nav3,nav4}. No instabilities were
observed in that case. However, the boundary conditions there were
computationally motivated rather than physically motivated. In
particular there was no obvious extension to asymptotically flat
spacetime.

In the present work we numerically explore semi-classical
evaporation in two spherically-reduced two-dimensional models of
asymptotically Minkowskian near-extremal black holes.  We find
results consistent with \cite{dibalowe}, indicating that
semi-classically there are no horizon instabilities during the
evaporation of charged black holes.

\section{The RN Model}
For our first model \cite{lowe,triv,trivstrom}, we begin with the
Einstein-Hilbert action
\be\label{einhilbaxn}
S=\frac{1}{2\pi}\int d^2x \sqrt{-g} e^{-2\phi}\left[R +
    2(\nabla\phi)^2 + 2\lambda^2 e^{2\phi}\right]~.
\ee
To this, we add an electromagnetic contribution:
\be\label{EMact}
S_{EM}=-\frac{1}{2\pi}\int d^2x \sqrt{-g}Q^2e^{2\phi}~,
\ee
where $Q$ is the charge of the black hole as it appears in the
spherically symmetric Reissner-Nordstr\"om metric,
\be\label{RNmet}
ds^2=-\left(1-\frac{2M}{r}+\frac{Q^2}{r^2}\right)dt^2 +
    \frac{1}{1-\frac{2M}{r}+\frac{Q^2}{r^2}}dr^2~.
\ee
The effective two dimensional action that describes (\ref{RNmet})
is given by
\be
S=\frac{1}{2\pi}\int d^2x \sqrt{-g} e^{-2\phi}\left[R +
    2(\nabla\phi)^2 + 2\lambda^2 e^{2\phi}-Q^2e^{4\phi}\right]~.
\ee
In order to study the dynamics, we add a large number $N$ of
matter fields, $f_i$,
\be\label{matterterm}
S_M=-\frac{1}{4\pi}\sum_{i=1}^{N}\int d^2x \sqrt{-g}(\nabla
    f_i)^2~.
\ee
This form of coupling is mostly justified by the ease of the
resulting solutions.  However, it also corresponds to the
bosonized version of the Callan-Rubakov modes of fermions coupled
to a magnetically charged black in four dimensions \cite{gidds}.

Additionally, we include the Liouville-Polyakov term
\be\label{polyakovterm}
S_{Q}=-\frac{N\hbar}{96\pi}\int d^2x \sqrt{-g} R  \Box^{-1} R ~,
\ee
which effectively accounts for the back-reaction of the matter
stress-energy tensor on the space-time geometry
\cite{polyakovpaper,polyakovbook}.

The full lagrangian in the conformal gauge becomes
\be\label{RNlagr}
{\cal L}=e^{-2\phi} \left(4\pt_+\pt_-\rho -
   4\pt_+\phi\pt_-\phi\right)+\lambda^2e^{2\rho}-Q^2e^{2\rho + 2\phi}
   -2\kappa \pt_+\rho \pt_-\rho ~.
\ee
>From this point on, we choose $\lambda=1$.
\subsection{Equations of Motion}
Variation of (\ref{RNlagr}) with respect to $\phi$ gives
\be
\pt_+\pt_-\rho - \pt_+\pt_-\phi + \pt_+\phi\pt_-\phi +
   \frac{Q^2}{4}e^{2\rho+4\phi}=0 ~.
\ee
A $\rho$ variation gives
\be
\frac{\kappa}{2}e^{2\phi}\pt_+\pt_-\rho - \pt_+\pt_-\phi + 2
   \pt_+\phi\pt_-\phi +
   \frac{e^{2\rho}}{4}(e^{2\phi}-Q^2e^{4\phi}) ~.
\ee
The resultant equations of motion can be written as
\bea
\label{rnrho}
\pt_+ \pt_- \rho &=& \frac{\pt_+ \phi \pt_- \phi + \frac{1}{4}
e^{2 \rho} (e^{2\phi} - 2 Q^2 e^{4 \phi})}{ 1 - \frac{\kappa} {2}
e^{2 \phi}} ~,\\
\label{rnphi}
\pt_+ \pt_- \phi &=& \pt_+ \pt_- \rho + \pt_+ \phi \pt_- \phi +
\frac{Q^2}{4} e^{2 \rho+4\phi} ~.
\eea
We may note that this model possesses a singularity for
$e^{-2\phi}=\kappa/2$.  In our solutions, this singularity will
lie safely behind the inner horizon.  It is useful
to rewrite these equations with
$\pht=e^{-\phi}$. Thus $\pt_+\pht=-\pt_+ \phi ~\pht$ and
$\pt_-\pt_+\pht=-\pt_-\pt_+ \phi ~\pht +\pt_-\phi \pt_+\phi
~\pht$. We can rewrite (\ref{rnphi}),
\be
\pt_+\pt_-\phi-\pt_+\phi \pt_-\phi =\frac{Q^2}{4}e^{2\rho+4\phi}+
  \frac{\pt_+\phi \pt_-\phi + \frac{e^{2\rho}}{4}(e^{2\phi} - 2 Q^2
  e^{4 \phi})}{ 1 - \frac{\kappa}{2}e^{2\phi}}~,
\ee
as
\be
-\frac{\pt_+\pt_-\pht}{\pht}=\frac{Q^2}{4}\frac{e^{2\rho}}{\pht^4}+
   \left(\frac{\pt_+\pht \pt_-\pht}{\pht^2} + \frac{e^{2\rho}}{4\pht^2}
   -\frac{Q^2e^{2\rho}}{2\pht^4}\right)/\left(1-\frac{\kappa}{2\pht^2}
   \right) ~.
\ee
Hence
\be
\pt_+\pt_-\pht=-\frac{Q^2}{4}\frac{e^{2\rho}}{\pht^3}-\frac{\pht\pt_+\pht
   \pt_-\pht+\frac{e^{2\rho} \pht}{4} - \frac{Q^2e^{2\rho}}{2\pht}}
   {\pht^2-\kappa/2} ~,
\ee
and,
\bea\label{RNrhophi}
\pt_+\pt_-\rho&=&\pt_+\pt_-\phi -
   \pt_+\phi \pt_-\phi - \frac{Q^2}{4}e^{2\rho+4\phi}\\
   &=&-\frac{\pt_+\pt_-\pht}{\pht}-\frac{Q^2e^{2\rho}}{4\pht^4}
   ~.
\eea
More concisely,
\bea\label{RNeomrho}
\pt_+\pt_-\rho&=&\frac{\pt_+\pht \pt_-\pht+\frac{e^2\rho}{4}\left(
   1-\frac{2Q^2}{\pht^2}\right)}{\pht^2-\kappa/2} ~,\\
\label{RNeomphi}
\pt_+\pt_-\pht&=&-(\pt_+\pt_-\rho)\pht-\frac{Q^2e^{2\rho}}{4\pht^3}
    ~.
\eea
\subsection{The Static Solution}
We first solve the equations of motion in the limit of the static
extremal black hole \cite{lowe}. (See also \cite{fabbri} for a recent
discussion of these static solutions, and \cite{lowecom} for analysis
of the full four-dimensional semi-classical equations of motion, in
the near-horizon limit.) We will later need these as
boundary conditions for the dynamic solutions. The model possesses
a linear dilaton vacuum solution corresponding to
\bea\label{RNLDV}
\phi &=& -\log\frac{1}{2}(x^+-x^-)~,\\
\rho &=& 0~.
\eea
This suggests to us a reasonable choice for the radial coordinate:
\be
\s = \frac{x^+-x^-}{2}~.
\ee
We note that
\be
\frac{\pt \s}{\pt x^+}= -\frac{\pt \s}{\pt x^-}=\frac{1}{2}
\ee
The equations of motion, (\ref{RNeomrho}-\ref{RNeomphi}), become
\bea\label{RNstatrho}
\rho'' &=& \frac{(\pht')^2 -e^{2\rho}\left(
   1-\frac{2Q^2}{\pht^2}\right)}{\pht^2-\kappa/2}~, \\
\label{RNstatpht}
\pht'' &=& -\rho''\pht + \frac{Q^2e^{2\rho}}{\pht^3}
    \;.
\eea
with the constraint that
\be
2\pht\left(\pht''-2\rho'\pht''\right)+\kappa\left(\rho''-(\rho')^2+t
    \right)=0~.
\ee
Linearizing about (\ref{RNLDV}), as $\s\rightarrow\infty$ we see
that
\bea
\s^2 \delta\rho'' &=& 1 + 2\delta\pht' -
    \left(1 + 2\delta\rho\right)\\
    &=&2 \delta\pht' + 2\delta\rho~,
\eea
and
\be
\delta\pht'' = -\s\delta\rho''~.
\ee
We can see that
\be
\delta \pht''=2\delta\rho'=-\s\delta\rho''
\ee
has the solution
\bea
\delta\rho' &=& \frac{M}{\s^2}~,\\
\delta\pht' &=& 1-\frac{2M}{\s}~.
\eea
As $\sigma \to \infty$ the linearized vacuum solutions are therefore
\bea\label{RNicrho}
\rho &=&-\frac{M}{\s}~,\\
\label{RNicphi}
\pht &=& \s - 2M \log \s~.
\eea
\subsection{The Boundary Conditions}
We will study black holes perturbed by shock waves of the matter
fields $f_i$. 
Our boundary conditions are chosen such that to the past of $x^+=0$, we have
the static extremal solution of (\ref{RNstatrho}),
(\ref{RNstatpht}) with the initial conditions at a chosen large
value of $\s=\s_{\infty}$ determined by (\ref{RNicrho}), (
\ref{RNicphi}). In practice, we make choices for $Q$ and $\kappa$,
and adjust $M$ so that a double horizon appears in the interior
region of the solutions.  Along a large value of
$x^-=x^-_{\infty}$, and to the future of $x^+=0$, 
we then impose the asymptotic linearized
solutions of a black hole with mass $M+\Delta M$
\bea
\pt_+\rho &=& 2\frac{M+\Delta M}{(x^+-x^-)^2}~,\\
\pt_+\pht &=& \frac{1}{2}-2\frac{M+\Delta M}{x^--x^-}~.
\eea
\section{The DW model}
The second model we consider, the so-called DW model, derives from
a general class of two-dimensional renormalizable generally
covariant field theories \cite{lowe,bol}
\be\label{genlag} {\cal
L}_{cl} = \sqrt{-g}(D(\phi) R + G(\phi)(\nabla\phi)^2 + H(\phi))~.
\ee
We require that the potentials in (\ref{genlag}) behave
asymptotically like those of the linear dilaton gravity model
considered above. That is,
\be
D(\phi) \longrightarrow \frac{G(\phi)}{4} \longrightarrow
\frac{H(\phi)}{4} \longrightarrow e^{-2\phi}
\ee
as $\phi \rightarrow -\infty$. After requiring, without loss of
generality, that $G(\phi) = -2 D'(\phi)$ and performing a
Brans-Dicke transformation on the metric $\hat{g} = e^{-2 \phi}
g$, the Lagrangian can be rewritten as
\be
\label{simp} \hat{{\cal
L}}_{cl} = \sqrt{-\hat{g}}\left(D(\phi)\hat{R} +
    W(\phi)\right)~,
\ee
where $W(\phi) = e^{2 \phi} H(\phi) $.  This is the form of the
lagrangian we refer to as the DW model. For these solutions, it is
convenient to work in the conformal gauge,
\be\label{confmet}
ds^2=-e^{2\rho}dx^+dx^-~.
\ee
Again, we add $N$ matter fields, (\ref{matterterm}), and a
back-reaction term, (\ref{polyakovterm}). Now
\bea\label{DWlag}
{\cal L}&=&{\cal L}_{cl} +\sum_{i=1}^N \pt_+ f_i \pt_- f_i -2
    \kappa\pt_+\rho\pt_-\rho \\
    &=&4D\pt_+\pt_-\rho+4D'\pt_+\phi\pt_-\phi+
    \frac{W}{2}e^{2\rho-2\phi}-2\kappa \pt_+\rho\pt_-\rho+\sum_{i=1}^N
    \pt_+ f_i \pt_- f_i ~.
\eea
where we have let $\kappa = N\hbar /12$.
\subsection{Equations of Motion}
Variation of (\ref{DWlag}) with respect to $\rho$ and $\phi$
yields
\be
4D''\pt_+\phi\pt_-\phi + 4D'\pt_+\pt_-\phi + We^{2\rho - 2\phi} +
4\kappa\pt_+\pt_-\rho = 0~,
\ee
and
\be
4D'\pt_+\pt_-\rho+4D''\pt_+\phi\pt_-\phi-8D'\pt_+\pt_-\phi -
8D''\pt_+\phi\pt_-\phi-(W-W'/2)e^{2\rho-2\phi} = 0
\ee
respectively.  Reexpressing these in a more useful form, we obtain
the equations of motion:
\bea\label{DWeom1}
\pt_+ \pt_- \phi &=& - \frac{D'' \pt_+\phi \pt_- \phi +
\frac{1}{4} e^{2 \rho- 2\phi} W +
\kappa \pt_+ \pt_- \rho} {D'}~, \\
\label{DWeom2}
\pt_+ \pt_- \rho &=& \frac{- D'' \pt_+ \phi \pt_- \phi -
\frac{1}{4} e^{2 \rho- 2\phi} (W+ \frac{W'}{2} )}{ D' + 2\kappa}~
.
\eea
As discussed in \cite{bol}, we must choose $D$ such that
$D'(\phi)<0$ for all $\phi$ in order to avoid a singularity of the
Brans-Dicke transformation. Also,
\be
\frac{d\phi}{d\s}\sim \int d\phi W(\phi)D'(\phi)
\ee
has zeroes along $\s$, the radial coordinate, equal to the number
of zeroes of $W(\phi)D'(\phi)$ plus one.  Therefore, for a charged
black hole, $W(\phi) D'(\phi)$ must possess a single zero, for
which $\phi(\s)$ will correspond to the extremal radius.  In our
solutions, we investigate the choice where $D= e^{-2 \phi} -
\gamma^2 e^{2 \phi}$ and $W= 4 - \mu^2 e^{4\phi}$. $\mu^2$, as the
coefficient of $e^{2 \phi}$, behaves as the charge of the black
hole.  This becomes clear through comparison with (\ref{EMact}),
from the Reissner-Nordstr\"om model.
\subsection{The Boundary Conditions}
The boundary conditions are set so that to the past of $x^+_0=1$ the
solution corresponds to the static, extremal solution of the
equations of motion, while along $x^-=-\infty$, and to the future of
$x^+_0=1$,  the boundary conditions are
\bea\label{shockdphi}
\phi = \rho &=& -\frac{1}{2}\log\left(M + \Delta M(x^+_0-x^+) -
    x^+x^-\right)~,\\
\pt_+ \phi = \pt_+ \rho &=& \frac{x^-+\Delta M}{2(M + \Delta
    Mx^+_0-x^+ x^- - \Delta M x^+ )} ~.
\eea
This is the same boundary condition imposed in \cite{cghs}. In
practice, the static solutions are solved with the static form of
(\ref{DWeom1}) and (\ref{DWeom2}), imposed at some large value of
$x^-=x^-_{\infty}$. Then, M is fine-tuned so that the there is a
double horizon static solution to the equations of motion,
(\ref{DWeom1}) and (\ref{DWeom2}). This solution in turn becomes
the boundary condition at $x^+_0=1$ for the full dynamical
solutions.
\section{Affine Coordinates}
We would like to answer questions regarding black hole stability
during the evaporation process.  An appropriate indicator would be
the behavior of the stress-energy as observed along a null
affinely parameterized geodesic.  The Christoffel symbols for the
metric (\ref{confmet}) are
\bea
\Gamma^+_{++}=2\pt_+ \rho ~,\\
\Gamma^-_{--}=2\pt_- \rho ~.
\eea
>From the geodesic equation for
$x^+$, we obtain
\be\label{geoeqnconf}
\frac{d^2 x^+}{d \xt+^2} +
2\pt_+ \rho\frac{dx^+} {d \xt^+}
    \frac{dx^+}{d \xt^+} = 0 ~,
\ee
with a similar equation for  $x^-$.  $\xt^+$ is an  affinely
parameterized null geodesic. If $x^-$ is held fixed at
$x^-=x^-_0$,
\be \pt_+ \rho\frac{d x^+}{d
\xt^+} = \frac{d
  \rho} {d \xt^+}~.
\ee
Then (\ref{geoeqnconf}) reduces to \be
\frac{d^2x^+}{d\xt^{+2}} + 2\frac{d \rho} {d \xt^+}\frac{dx^+}{d
    \xt^+}=0~.
\ee
Thus,
\be
\frac{d}{d\xt^+}\left(\log\frac{dx^+}{d\xt^+}\right)=-2\frac{d\rho}{d\xt^+}~,
\ee
which is then solved to give
\be\label{cchangep} \frac{d\xt^+}{d
x^+}= C(x^-_0) e^{2\rho(x^+,x^-_0)}~,
\ee
where $C(x^-)$ is a constant of integration which we will soon
determine.  Similarly, at fixed $x^+=x^+_0$,
\be\label{cchangem}
\frac{d\xt^-}{d x^-}= C(x^+_0) e^{2\rho(x^+_0,x^-)}~.
\ee
Thus we
make calculations of $\tilde{T}$, as defined by
\be\label{afstresseqns} \tilde{T}_{\pm\pm}\equiv
    \left(\frac{d\xt^{\pm}}{dx^{\pm}}\right)^2T_{\pm\pm}=
    \frac{C(x^{\mp}_0)}{e^{4\rho(x^{\pm})}}T_{\pm\pm}~.
\ee
This gives us an observable measure of the energy flux that is
independent of the choice in coordinates.

We choose the normalization constants,
$C(x^-_0)$, $C(x^+_0)$ in (\ref{cchangep}) and (\ref{cchangem}), so
the affine coordinates coincide with asymptotically Minkowskian
coordinates far from the black hole.
This
normalization will not be necessary in the RN case, because there the
coordinates are already asymptotically Minkowskian.  
For the DW model, we choose $C(x^-_0)$, $C(x^+_0)$ with the requirement
that along some large chosen value of $x^+$, termed
$x^+=x^+_{\infty}$, and along a similarly chosen
$x^-=x^-_{\infty}$, \be ds^2=-d\xt^+d\xt^-~,
\ee so that, \be
e^{2\rho}\frac{dx^+}{d\xt^+}\frac{dx^-}{d\xt^-}=1~.
\ee Given that asymptotically $e^{-2\rho}=M-x^+x^-$, then this
amounts to requiring that at $x^-=x^-_0$ \be
\frac{dx^+}{d\xt^+}=e^{-2\rho(x^+,x^-_0)}\frac{\sqrt{M -
    x^+_{\infty}x^-_{\infty}}}{M - x^+_{\infty}x^-_0}~,
\ee and at $x^+=x^+_0$ \be
\frac{dx^-}{d\xt^-}(x^+_0)=e^{-2\rho(x^+_0,x^-)}\frac{\sqrt{M -
    x^+_{\infty}x^-_{\infty}}}{M - x^+_0x^-_{\infty}}~.
\ee Using these we obtain meaningfully normalized null affine
coordinates with which to calculate the stress-energy tensors of
(\ref{afstresseqns}).
\section{Results}
Careful analysis limits the range of values of our variables,
($\gamma$, $\mu$ and $\kappa$, in the DW model, $Q$ and $\kappa$
in the RN) for which numerical errors are sufficiently small to
permit unambiguous statements. We compared our results against the
classical solutions to gauge the buildup of numerical error. 
These numerical constraints prevent us from making the back-reaction
arbitrarily weak. A useful measure of the strength of the
back-reaction
is the ratio of the time-scale of the quantum evaporation to the
light-crossing time of the black hole. These time-scales may be
extracted by inspection of the equations of motion, with the results 
that  $t_{light}/t_{evap} = \kappa/Q^2$ for the RN model, and 
$t_{light}/t_{evap} = \kappa/\mu$ for the DW model.  
We quote the results for the regime when this ratio is
of order one, where numerical errors are negilible. The qualitative
behavior of the solutions does not change as $\kappa$ is made smaller, 
as far as we have been able to determine.
Of course, the
semi-classical approximation to the 
evaporation is no less valid when we do not impose the weak
back-reaction condition.

\begin{figure}[htbp] \includegraphics[width=3.5in]{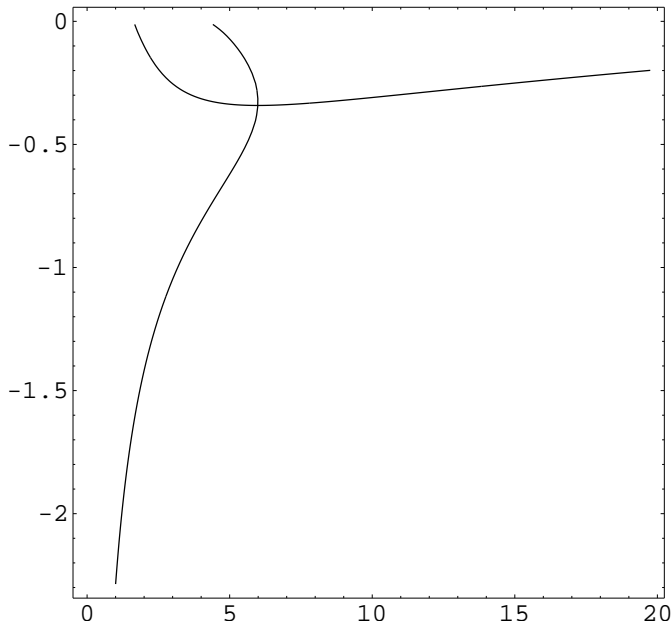}
\caption{{\bf DW model:} The outer horizon is shown as it recedes
to meet
  the inner horizon, in turn moving out, at $\phi_0$, the extremal
  radius. $x^+$ and $x^-$ are the $x$ and $y$ axes respectively.
  [$\gamma=8; \mu=15; \kappa=10; \Delta M=1.5$]}
\label{DWcontours} \end{figure}
\begin{figure}[htbp]
\includegraphics[width=3.5in]{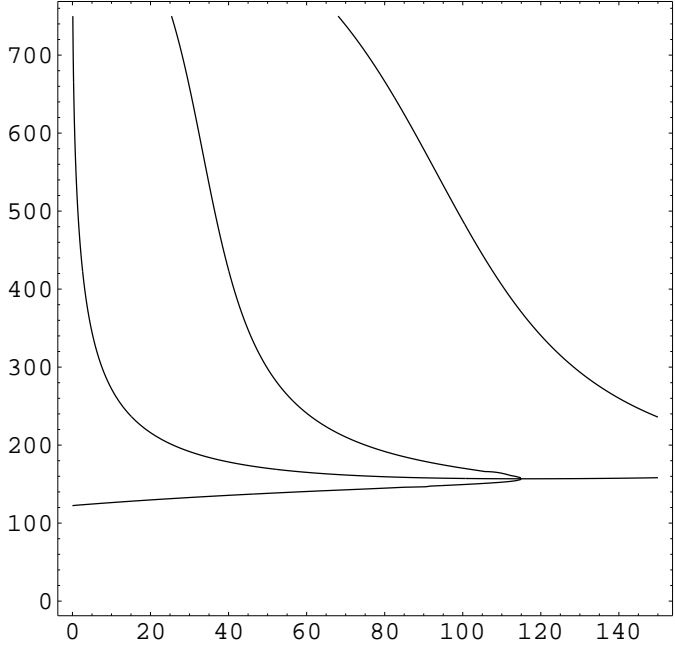}
\caption{{\bf RN model:} The outer horizon is shown as it recedes
to meet
  the inner horizon, in turn moving out, at $\pht_0$, the extremal
  radius. $x^+$ and $x^-$ are the $x$ and $y$ axes respectively.
  The additional contour lying above these is a remnant of
  $\pht_0$.  No particular meaning has been attributed to this as it
  lies in a causally inaccessible region.
  [$Q=\sqrt{60}; \kappa=40; \Delta M=.2$]}
\label{RNcontours}
\end{figure}
\begin{figure}[htbp] \includegraphics[width=3.5in]{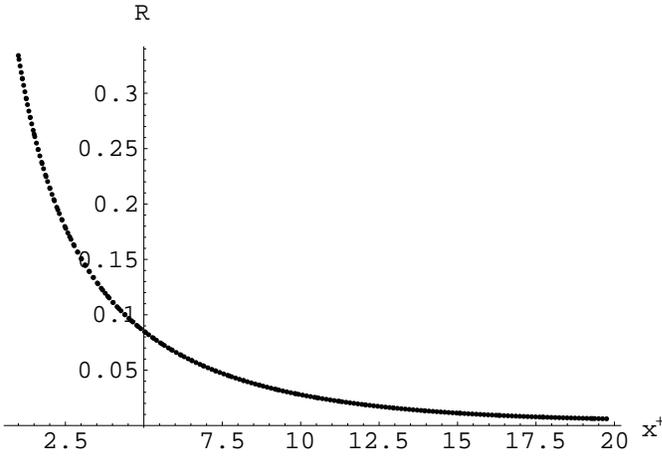}
\caption{{\bf DW model:} The Ricci scalar curvature, $R$,
calculated along the contour $\phi=-1.3$ throughout the
evaporation process.  The asymptotic value to which $R$ settles
coincides with its extremal value. [$\gamma=8; \mu=15; \kappa=10;
\Delta M=1.5$]}
\label{DWcurv} \end{figure}
\begin{figure}[htbp]
\includegraphics[width=3.5in]{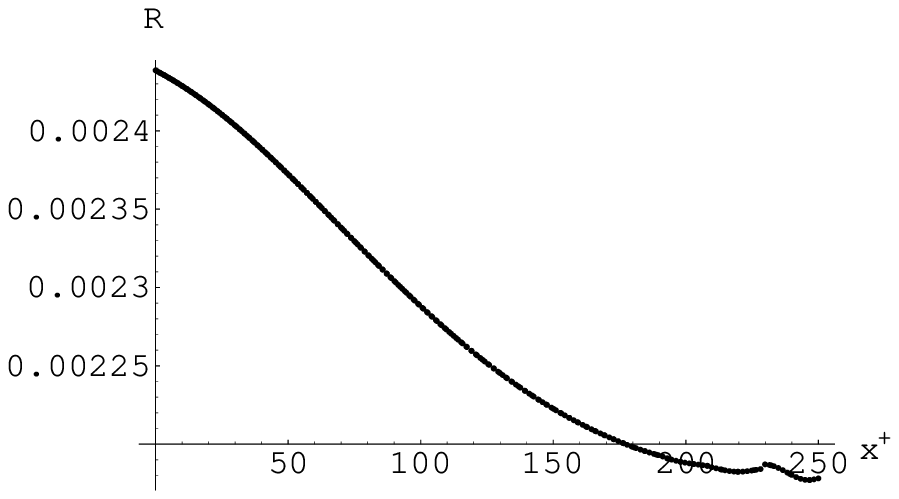}
\caption{{\bf RN model:} The Ricci scalar curvature, $R$,
calculated along the contour $\pht=15$ throughout the evaporation
process.  The asymptotic value to which $R$ settles coincides with
its extremal value. [$Q=\sqrt{60}; \kappa=40; \Delta M=.2$]}
\label{RNcurv}
\end{figure}

A typical contour plot is shown for each model in
FIG.~\ref{DWcontours} and FIG.~\ref{RNcontours}. As a shock mass
is sent in, the two apparent horizons, given by the zeroes of $\pt_+\phi$,
split apart, with the outer apparent horizon moving past $\phi_0$,
while the inner horizon moves behind this radius. Immediately
afterwards, the outer horizon begins to recede and meets the inner
horizon back at the extremal radius $\phi_0$ (the value of $\phi$ at
the position of the horizon of the initial extremal black hole). 

It is useful to regard the separation of the horizons along the $x^-$
direction as a measure of the excitation energy of the extremal black hole
(this statement can be made more precise in the adiabatic
approximation \cite{trivstrom,lowe}). This indicates
the semi-classical solutions break
down before the meeting point of the apparent horizons
 since here the energy of an
emitted quanta inevitably becomes comparable to the energy of
excitation above the extremal ground state \cite{wilczetc}.
Evolving for points in the causal future of this endpoint would no
longer be consistent with the semi-classical approximation.

As discussed in
\cite{lowe}, the Ricci scalar curvature calculated along a given
contour of $\phi$ (see FIG.~\ref{DWcurv} and FIG.~\ref{RNcurv})
demonstrates a return to the extremal quantity as evaporation
proceeds, strongly indicating a smooth evaporation process.  This
is to say that an observer at a fixed radius outside of the black
hole sees the same physical environment towards the endpoint of
evaporation as was seen before the shock mass was introduced.
\begin{figure}[htbp] \includegraphics[width=3.5in]{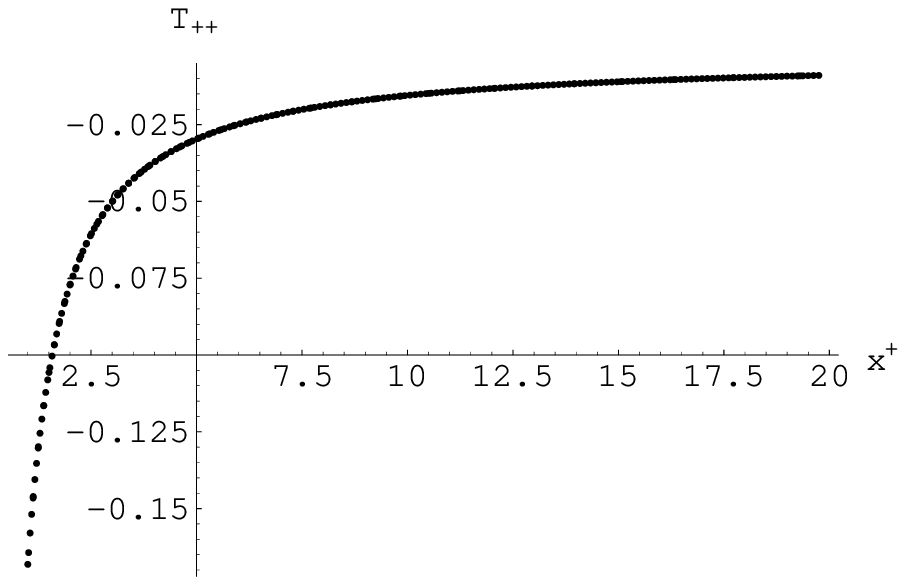}
\caption{{\bf DW model:} $\tilde{T}_{++}$ calculated along the
contour $\phi=-1.3$ varies smoothly throughout the evaporation
process.  The data suggests that the flux $\tilde{T}_{++}$
approaches zero, or a small constant, as the shock mass is
evaporated away. [$\gamma=8; \mu=15; \kappa=10; \Delta M=1.5$]}
\label{DWtppplot} \end{figure}
\begin{figure}[htbp] \includegraphics[width=3.5in]{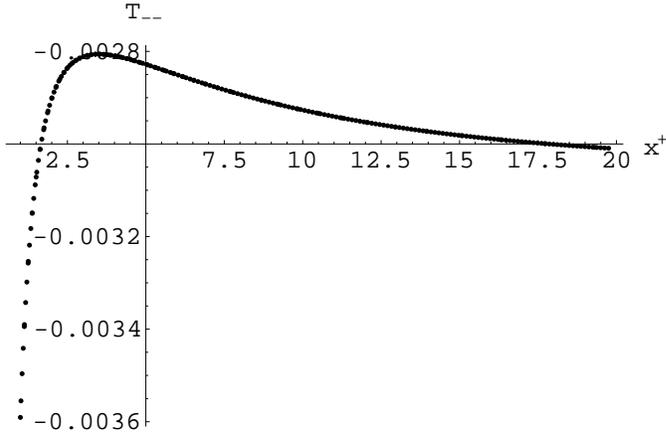}
\caption{{\bf DW model:} $\tilde{T}_{--}$ calculated along the
contour $\phi=-1.3$ varies smoothly throughout the evaporation
process. As the endpoint of evaporation is reached, the data
suggests that this quantity approaches some constant value.
[$\gamma=8; \mu=15; \kappa=10; \Delta M=1.5$]} \label{DWtmmplot}
\end{figure}
\begin{figure}[htbp] \includegraphics[width=3.5in]{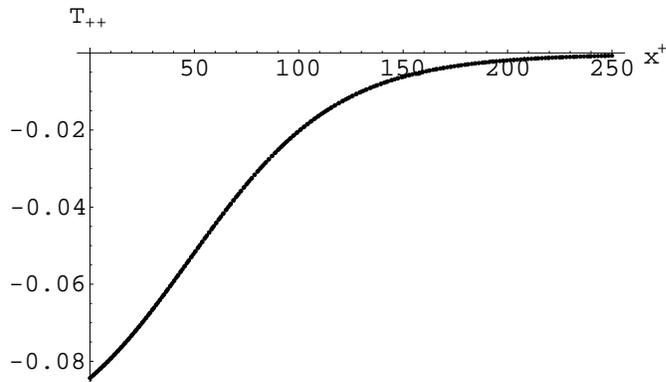}
\caption{{\bf RN model:} $\tilde{T}_{++}$ calculated along the
contour $\pht=15$ throughout the evaporation process.  After the
shock mass is introduced at $x^+_0$, the flux $\tilde{T}_{++}$,
which is negative, effectively approaches zero. [$Q=\sqrt{60};
\kappa=40; M=5.641; \Delta M=.2$]} \label{RNtppplot} \end{figure}
\begin{figure}[htbp] \includegraphics[width=3.5in]{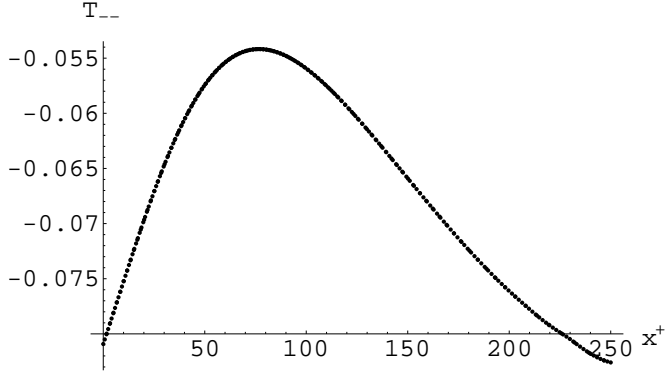}
\caption{{\bf RN model:} $\tilde{T}_{--}$ calculated along the
contour $\pht=15$ throughout the evaporation process.
[$Q=\sqrt{60}; \kappa=40; \Delta M=.2$]} \label{RNtmmplot}
\end{figure}
\begin{figure}[htbp] \includegraphics[width=3.5in]{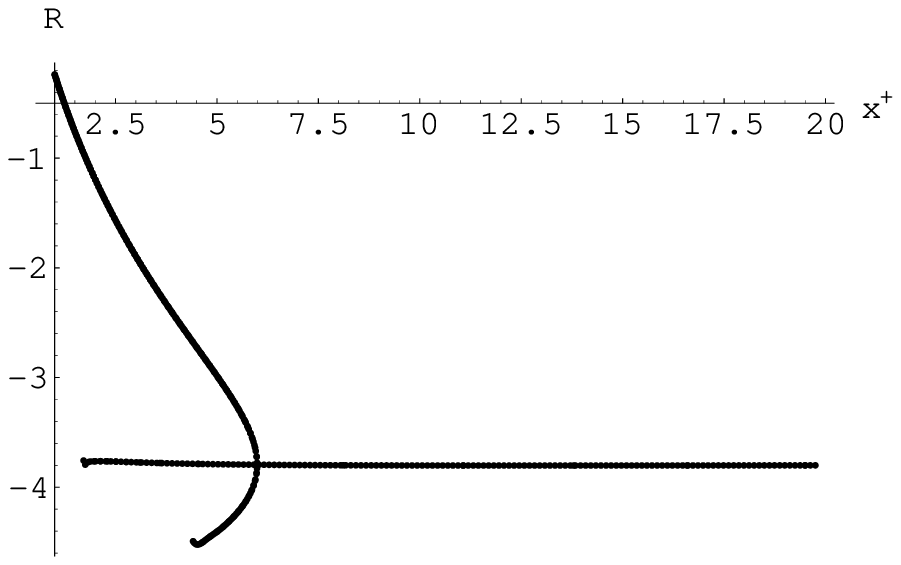}
\caption{{\bf DW model:} The Ricci scalar curvature, $R$,
evaluated at points along the outer horizon, the inner horizon,
and $\phi_0$, overlapped.  The absolute value of the 
curvature at the outer horizon is
increasing, while decreasing at the inner horizon.  When the
horizons meet again at the extremal radius, $\phi_0$, the
curvature is at its extremal value, indicating that the horizon
environment has returned to its original state. [$\gamma=8;
\mu=15; \kappa=10; \Delta M=1.5$]} \label{DWhorcurv} \end{figure}
\begin{figure}[htbp]
\includegraphics[width=3.5in]{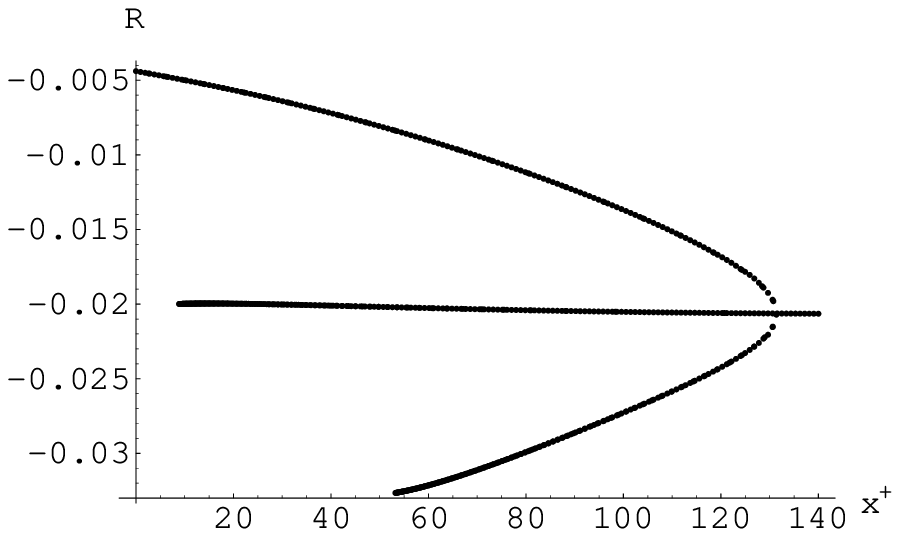}
\caption{{\bf RN model:} The Ricci scalar curvature, $R$,
evaluated at points along the outer horizon, the inner horizon,
and $\pht_0$, overlapped.  The absolute value of the 
curvature at the outer horizon is
increasing, while decreasing at the inner horizon.  When the
horizons meet again at the extremal radius, $\pht_0$, the
curvature is at its extremal value, indicating that the horizon
environment has returned to its original state. [$Q=\sqrt{60};
\kappa=40; \Delta M=.2$]}
\label{RNhorcurv}
\end{figure}

For the DW model, we also evaluate $\tilde{T}_{\pm\pm}$ (as shown
in FIG.~\ref{DWtppplot} and FIG.~\ref{DWtmmplot}), calculated from
(\ref{afstresseqns}) along the fixed $\phi$ contour in FIG.~
\ref{DWcontours}. The behavior of these quantities is similarly
consistent with a smooth evaporation process, returning the
near-extremal black hole back to the extremal ground state. These
fluxes are non-zero initially, but approach zero during
evaporation.  In the RN case, however, the flux perceived by an
observer at a fixed radius outside of the black hole does not go
to zero. This is evident from inspecting the behavior of $T_{--}$,
shown in FIG.~\ref{RNtmmplot}, whose absolute value does not
decrease monotonically, indeed eventually rising again. This is a
not an entirely unexpected result, since the affine coordinates are
not true asymptotically Minkowskian coordinates, so the flux need not
vanish at infinity. $T_{++}$ (FIG.~\ref{RNtppplot}), however, does approach a
zero value with increasing $x^+$. 

\begin{figure}[htbp] \includegraphics[width=3.5in]{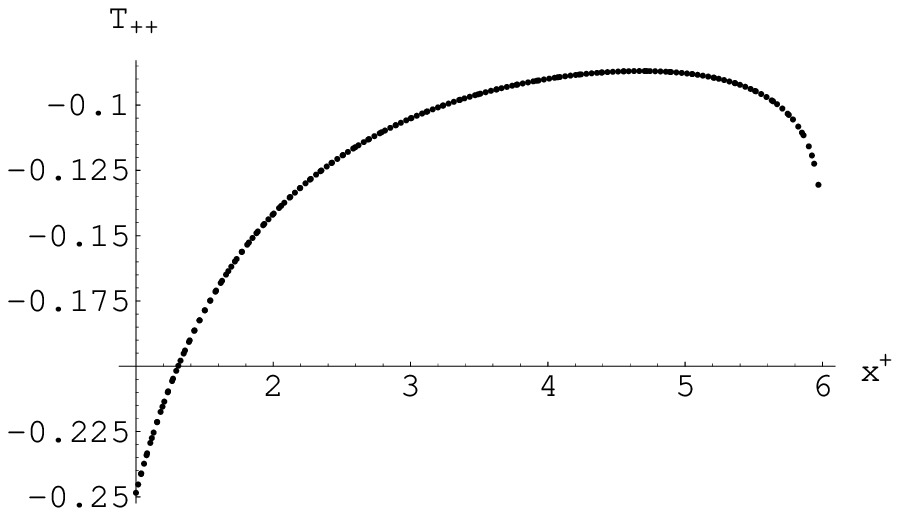}
\caption{{\bf DW model:} $\tilde{T}_{++}$ evaluated at points
along the outer
    horizon, shown as it varies smoothly for increasing $x^+$.
  [$\gamma=8; \mu=15; \kappa=10; \Delta M=1.5$]}
\label{DWoutertpp} \end{figure}

\begin{figure}[htbp] \includegraphics[width=3.5in]{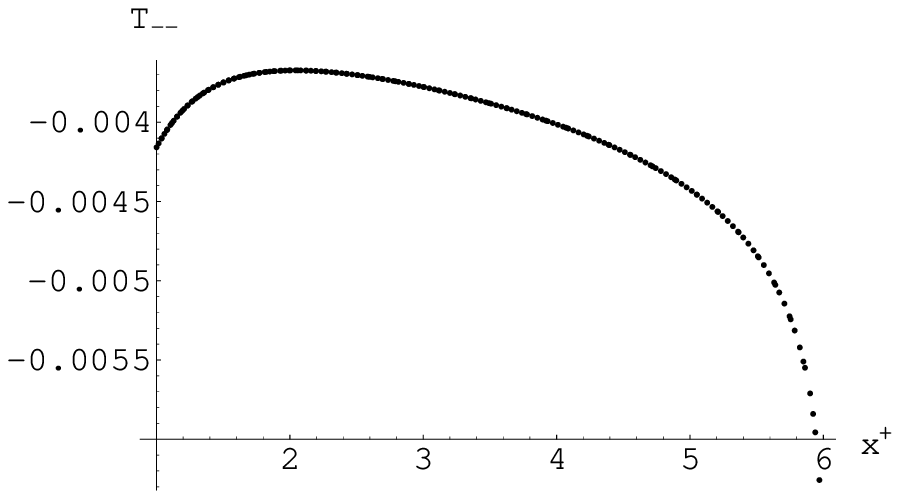}
\caption{{\bf DW model:} $\tilde{T}_{--}$ evaluated at points
along the outer
    horizon, shown as it varies smoothly for increasing $x^+$.
    [$\gamma=8; \mu=15; \kappa=10; \Delta M=1.5$]}
\label{DWoutertmm} \end{figure}

\begin{figure}[htbp] \includegraphics[width=3.5in]{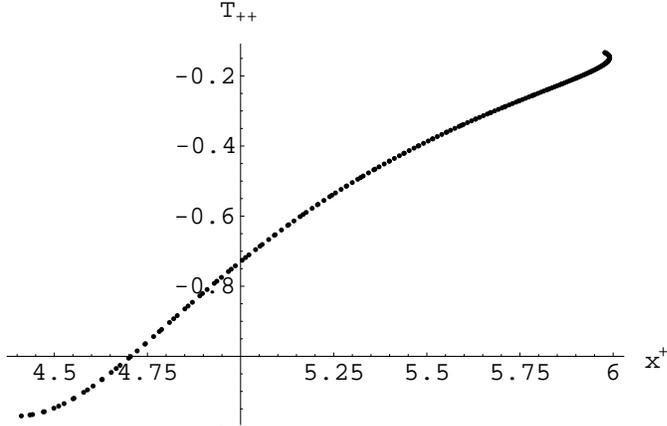}
\caption{{\bf DW model:} $\tilde{T}_{++}$ evaluated at points
along the inner
    horizon, shown as its value decreases smoothly for increasing $x^+$.
  [$\gamma=8; \mu=15; \kappa=10; \Delta M=1.5$]}
\label{DWinnertpp} \end{figure}

\begin{figure}[htbp] \includegraphics[width=3.5in]{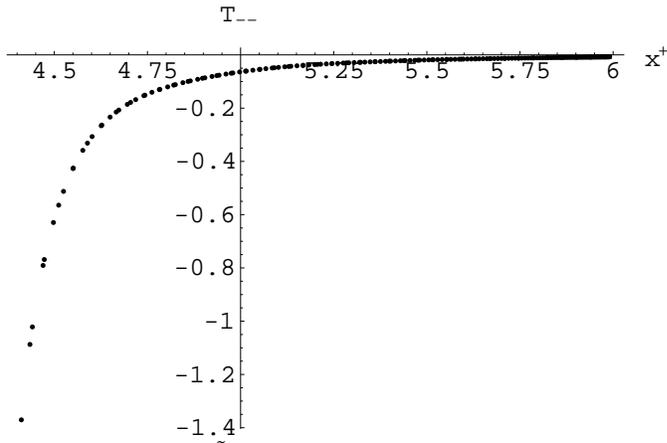}
\caption{{\bf DW model:} $\tilde{T}_{--}$ evaluated at points
along the inner
    horizon, shown as it varies for increasing $x^+$.  This flux
    goes to zero as the shock mass evaporates away.
  [$\gamma=8; \mu=15; \kappa=10; \Delta M=1.5$]}
\label{DWinnertmm} \end{figure}

\begin{figure}[htbp]
\includegraphics[width=3.5in]{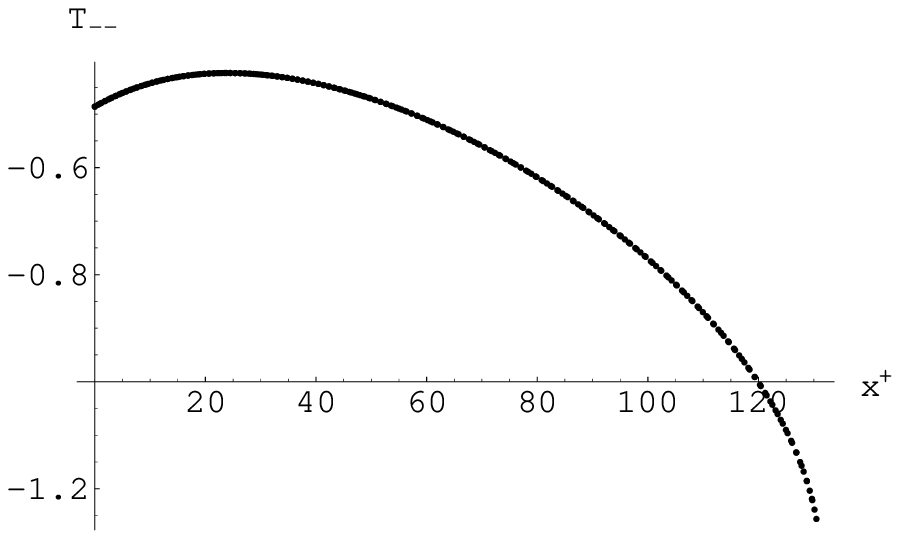}
\caption{{\bf RN model:} $\tilde{T}_{--}$ evaluated at points
along the outer
    horizon, shown as it varies smoothly for increasing $x^+$.
    [$Q=\sqrt{60}; \kappa=40; \Delta M=.2$]}
\label{RNoutertmm}
\end{figure}

\begin{figure}[htbp]
\includegraphics[width=3.5in]{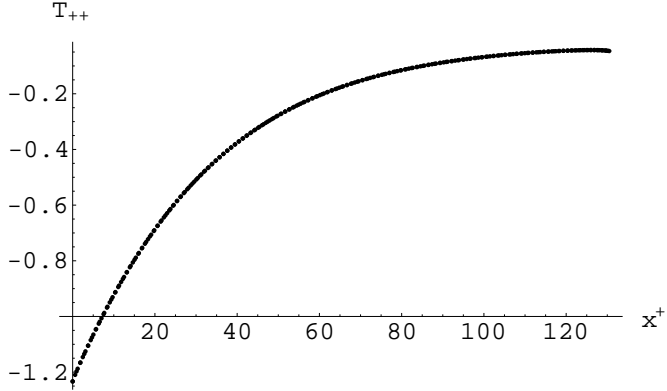}
\caption{{\bf RN model:} $\tilde{T}_{++}$ evaluated at points
along the outer
    horizon, shown as it varies smoothly for increasing $x^+$.
    The $\tilde{T}_{++}$ flux through the outer horizon appears to
    vanish as evaporation proceeds.
  [$Q=\sqrt{60}; \kappa=40; \Delta M=.2$]}
\label{RNoutertpp}
\end{figure}

\begin{figure}[htbp]
\includegraphics[width=3.5in]{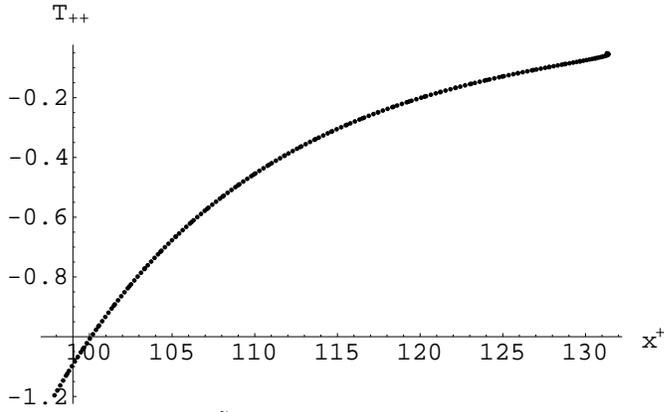}
\caption{{\bf RN model:} $\tilde{T}_{++}$ evaluated at points
along the inner
    horizon, shown as it varies for increasing $x^+$.  This flux
    decreases in absolute value throughout the evaporation
    process.  [$Q=\sqrt{60}; \kappa=40; \Delta M=.2$]}
\label{RNinnertpp}
\end{figure}

\begin{figure}[htbp]
\includegraphics[width=3.5in]{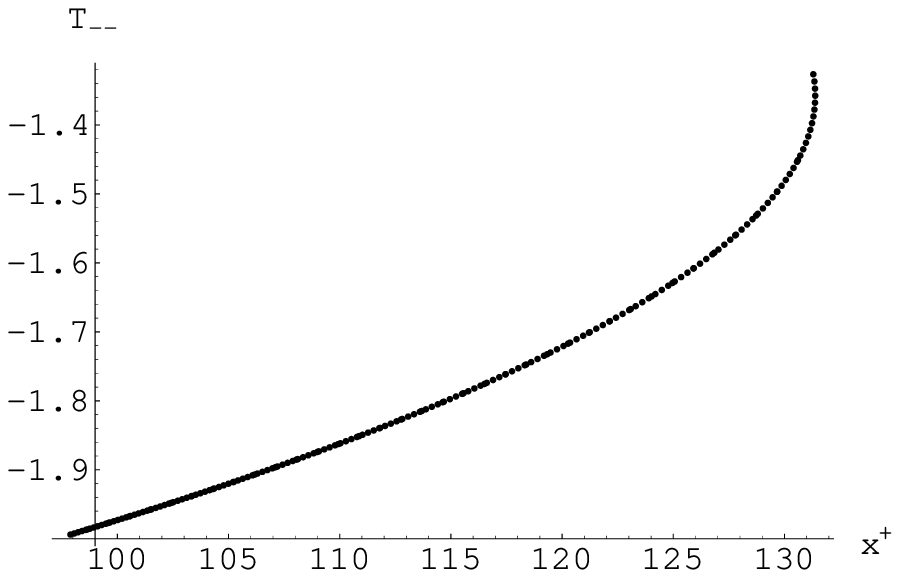}
\caption{{\bf RN model:} $\tilde{T}_{--}$ evaluated at points
along the inner
    horizon, shown as it varies smoothly for increasing $x^+$.
  [$Q=\sqrt{60}; \kappa=40; \Delta M=.2$]}
\label{RNinnertmm}
\end{figure}

In order to directly test for instabilities in the inner or outer
apparent horizon we also inspect the behavior of the stress-energy momentum
and the Ricci scalar curvature for the contours of $\pt_+\phi=0$
along the outer and inner apparent horizons in order to see if an
instability arises. FIG.~\ref{DWhorcurv} and FIG.~\ref{RNhorcurv}
appear to indicate that the scalar curvature evolves smoothly
throughout evaporation and returns to its extremal value.
Meanwhile, FIG's.~\ref{DWoutertpp}~-~\ref{RNinnertmm} demonstrate
that for both models the absolute flux through the horizons
decreases during the evaporation, consistent with a decreasing
black hole mass.  More importantly, there is no indication of a
buildup of energy behind either the inner or outer apparent
horizon.  The smooth variation of $T_{++}$ and $T_{--}$ in these
regions suggests a return, through evaporation, to the extremal
ground state after the initial excitation.


\section{Discussion}
The pictures emerging from both the DW and RN case were
consistent.  After the injection of shock matter, the
near-extremal black hole evaporates and returns to the extremal
ground state. The outer and inner horizon experience no
distressing energy instabilities and come back together, rejoining
at the extremal radius. Furthermore, studies of the Ricci scalar
and the stress-energy tensors calculated along null affine
geodesics, indicate that for observers outside the black hole, the
environment returns to its original state.

The original motivation for studying the evaporation of charged
black holes was to shed light on the black hole information
puzzle. Quantum mechanics necessarily requires unitarity. However,
semi-classically, the emitted radiation from black holes possesses
a thermal distribution.  This leads to the information puzzle:
what happens to the information contained in the ingoing matter
once the mass of the black hole is radiated away?

While it may be that physics in the real universe is simply not
unitary \cite{hawk1,hawk2}, other resolutions have nevertheless
been suggested. One such potential resolution of the paradox is
that the information of the ingoing matter becomes stored in
stable remnants that remain at the end of evaporation. These
remnants would apparently need to contain an arbitrarily large
amount of information despite possessing a Planck-sized mass
\cite{preskill}. In this regard, extremal black holes hold some
promise; behind the charged black hole horizon lies an infinite
throat region \cite{bols} that can apparently store an arbitrary
amount of information. This fact is also manifested by the
infinite degeneracy of the extremal ground state \cite{trivstrom}
at the semi-classical level. Thus extremal black holes represent
the testing ground for the remnant resolution to the information
paradox \cite{gidinfo1,gidinfo2}. There is some debate as to
whether or not the pair production of such charged black holes
occurs at a finite rate \cite{bols,gidinfo1,sussk,gidinfo2}, given
the infinite degeneracy of the ground state.  However, this
infinite degeneracy is likely a result of the semiclassical
averaging over the quantum geometries.  This should not occur in a
full quantum treatment. For example, D-brane calculations
demonstrate that summing over microstates in fact yields a finite
Bekenstein-Hawking entropy \cite{stromvafa}. Matrix theory also
indicates a similar result \cite{lowematrix}. This may lay to rest
infinite production concerns. Our studies did not reveal an
instability in the evolution of excited extremal black holes. Thus
as it stands, we cannot yet rule them out as the ultimate devices
for information storage.

Another suggested resolution to the information puzzle proposes
that higher order quantum corrections would reveal that the
outgoing radiation is not strictly thermal, carrying off
information about the ingoing radiation in the form of quantum
correlations with the ingoing matter.  This, however, while at
first seeming to be a most agreeable resolution, presents other
theoretical difficulties. Once matter disappears behind the
horizon, it becomes causally dissociated from the outside region.
Therefore such cross-horizon correlations would necessarily
violate causality, unless some sort of information stripping were
to occur at the horizon itself.  At the semi-classical level,
however, we do not expect anything special to occur at the horizon
in the frame of reference of the in-falling observer.  This issue
can be resolved by appealing to a ``stretched horizon'' membrane
picture \cite{bhcomp}, based on the principle of black hole
complementarity.  This principle asserts that matter which has
fallen past the event horizon and the Hawking radiation emitted
from there are not different objects; they are equivalent and
complementary descriptions from reference frames which can be
related to each other by large Lorentz boosts.  Since there is no
communication to outside of the black hole once an observer has
passed in, there is no logical contradiction by prescribing that
unitarity remains valid entirely outside of the horizon.
Furthermore, the contribution of non-local terms arising from
string theory \cite{justlowe,loweetc} may account for
cross-horizon correlations which will nevertheless be undetectable
by any single observer.  That is, if we do not require a local
field theory, except at low energy, it becomes possible to
construct a fully self-consistent causal theory where field
operators behind and in front of the horizon will not commute
\cite{LSU}.  String theory and the existence of quantum
correlations between the ingoing and outgoing matter then
ultimately promises to contain the key for resolving the
information puzzle.

The models we have considered here suggest that at the
semi-classical level, upon entering the black hole, information is
lost to the outside world.  However, while it becomes
inaccessible, the information enters the infinite asymptotic
throat region behind the horizon and is not actually destroyed.
While this picture permits remnants to contain the ``lost''
information, it still remains possible that the non-local string
theory terms do indeed represent the real universe and the
information does return to the outside region.  Our success lies
in showing that the semi-classical treatment of charged black hole
evaporation does not breakdown sooner than would be expected from
\cite{wilczetc}, despite heuristic arguments of Jacobson
\cite{jacobson} to the contrary.  Results presented here are also
consistent with analytical results found in \cite{dibalowe}.

\end{document}